# UNIFIED MODEL FOR P-N JUNCTION CURRENT-VOLTAGE CHARACTERISTICS

**M. J. Cristea**

'Politehnica' University of Bucharest, Faculty of Electronics, Telecom. and Information Technol.
E-mail: mironmail@gmail.com URL: http://arh.pub.ro/mcristea

*Abstract*
*The current-voltage p-n junction characteristics were mainly analyzed until now at low injection levels and high level injection separately. This work unifies the low injection, medium injection, high injection levels and the ohmic region of the I/V characteristic*
*Keywords: p-n junctions, current-voltage characteristics, low injection and high injection levels, ohmic region.*

## 1. INTRODUCTION

The p-n junction current-voltage characteristics have been studied from the beginnings of semiconductor based electronics, to remind only the ideal diode equation of W. Shockley [1], giving the voltage dependence of the diffusion current at low injection levels. Since then, the recombination component current-voltage dependence was determined and also the high injection level area has been investigated, although in a lesser amount, mainly concerning p-i-n diodes and bipolar transistors. A unified theoretical model for the p-n junction characteristics for both low and high levels of injection is still missing and the used equation involving the $\theta$ parameter is empirical and derived from the Ebers-Moll model of the bipolar transistor [2].

## 2. THEORY

While the low level injection region of the p-n junction characterictics is well studied and with an undisputed equation originating from Shockley, the high level area is more tributary to experimental measurements and empirical formulas and the theoretical approaches are often affected by errors, like for example the application of the well known Shockley boundary condition at high injection levels. The correct boundary conditions followed by the integration of the semiconductor equations will be next carried out next.

### 2.1. THE BOUNDARY CONDITIONS FOR P-N SEMICONDUCTOR JUNCTIONS AT MEDIUM AND HIGH INJECTION LEVELS

In a non-degenerated semiconductor material, the concentrations of electric charge carriers are given by:

$$n = N_c \cdot \exp\left(\frac{E_{fn} - E_c}{kT}\right) \quad (1)$$

$$p = N_v \cdot \exp\left(\frac{E_v - E_{fp}}{kT}\right) \quad (2)$$

with $n$ and $p$ the concentration of electrons and holes, $N_c$ and $N_v$ the densities of states in the conduction and valence bands, $E_c$ and $E_v$ the limits of the conduction and valence energy bands and $E_{fn}$ and $E_{fp}$ the quasi-Fermi energy levels in the space-charge region of the p-n junction for electrons and for holes, respectively. $k$ is the Boltzmann constant and $T$ is the absolute temperature.

The difference between the quasi-Fermi energy levels, normalized to the electric charge of the electron $q$ equals the bias voltage across the p-n junction is:

$$V_J = \frac{E_{fn} - E_{fp}}{q} \quad (3)$$

Therefore, the $pn$ product in the depletion region of the p-n junction is given by:

$$pn = n_i^2 \exp\left(\frac{qV_J}{kT}\right) \quad (4)$$

with $n_i$ the intrinsic carrier concentration of the semiconductor.

Using (4), the minority carrier concentrations at the boundaries of the space-charge region are calculated:

$$p_n = \frac{n_i^2}{n_n} \exp\left(\frac{qV_J}{kT}\right) \quad (5a)$$

$$n_p = \frac{n_i^2}{p_p}\exp\left(\frac{qV_J}{kT}\right) \quad (5b)$$

At **low injection levels**, the majority carrier concentrations remain equal with the doping levels $N_D$ (donors) and $N_A$ (acceptors), therefore:

$$p_n = \frac{n_i^2}{N_D}\exp\left(\frac{qV_J}{kT}\right) \quad (6a)$$

$$n_p = \frac{n_i^2}{N_A}\exp\left(\frac{qV_J}{kT}\right) \quad (6b)$$

By taking into account the expressions of the minority carrier concentrations at equilibrium $p_{n0}$ and $n_{p0}$, the classical Shockley formulas [1] are obtained:

$$p_n = p_{n0}\exp\left(\frac{qV_J}{kT}\right) \quad (7a)$$

$$n_p = n_{p0}\exp\left(\frac{qV_J}{kT}\right) \quad (7b)$$

At **high current injection levels**, both the minority and majority carrier concentrations surpass the doping level of the semiconductor, and due to the charge equilibrium condition, they are equal in magnitude. For example, in a p-n⁻ junction, where the high injection level occurs in the lightly doped n⁻ side of the junction:

$$p_n = n_n \gg n_{n0} = N_D \quad (8)$$

Therefore, using (8) and replacing $n_n$ with $p_n$ in equation (5a), the new boundary condition for the n side of the p-n junction at high injection levels is obtained:

$$p_n(x_n) = n_i \exp\left(\frac{qV_J}{2kT}\right) \quad (9)$$

The formula of $n_p(x_p)$ in the p side of the junction is the same, since it doesn't depends on the doping levels. ($x_n$ and $x_p$ are the space charge region boundaries.)

At **medium current injection levels**, the simplifying conditions $n_n = n_{n0} = N_D$ - low injection level or $n_n = p_n \gg n_{n0}$ - high injection level (p-n⁻ junction) are not valid, therefore equations (5a) and (5b) must be calculated in the general case.

To do this, equation (5a) must be used in conjunction with the electric charge equilibrium equation:

$$p_n - n_n + N_D = 0 \quad (14)$$

The following expression is obtained for the minority carrier concentration $p_n$ at the boundary of the space charge region in the n side of the junction:

$$p_n(x_n) = \sqrt{n_i^2 \exp\left(\frac{qV_J}{kT}\right) + \frac{N_D^2}{4}} - \frac{N_D}{2} \quad (15)$$

It is easy to notice that this equation reduces to equation (9) when the terms including the doping levels are negligible – high injection level conditions. It also reduces to the classical Shockley condition in the case of low level injection, and this can be obtained by taking $N_D$ as common factor in equation (15), expanding the square root into series and retaining only the first two terms:

$$p_n = \frac{N_D}{2}\left(\sqrt{1 + \frac{4n_i^2}{N_D^2}\exp\left(\frac{qV_J}{kT}\right)} - 1\right) \quad (16)$$

In the case of low injection level, the second term of the square root is much smaller than unity; therefore the square root can be expanded into series and only first two terms retained:

$$p_n = \frac{N_D}{2}\left[1 + \frac{2n_i^2}{N_D^2}\exp\left(\frac{qV_J}{kT}\right) - 1\right] =$$
$$= \frac{n_i^2}{N_D}\exp\left(\frac{qV_J}{kT}\right) \quad (17)$$

Thus, equation (15) represents the general formula for the boundary conditions of p-n junctions, irrespective of the injection levels at which those junctions are operated.

## 2.2. THE INTEGRATION OF THE SEMICONDUCTOR EQUATIONS

The low injection case means the integration of the diffusion equation for holes (p-n⁻ junction)

$$D_p \frac{d^2 p_n}{dx^2} - \frac{p_n - p_{n0}}{\tau_p} = 0 \quad (18)$$

with equation (9) as boundary condition at the edge of the space charge region in the n⁻ side of the junction. The holes concentration is obtained as a well-known exponential dependence with the decay constant $L_p = \sqrt{D_p \tau_p}$ - the holes diffusion length.

$$p_n(x) = p_{n0} + p_{n0}\left[\exp\left(\frac{qV_J}{kT}\right) - 1\right]\exp\left(-\frac{x - x_n}{L_p}\right) \quad (19)$$

The current equation is then:

$$J = -qD_p \frac{dp_n}{dx}\bigg|_{X=X_n} = \frac{qD_p}{L_p} p_n(x_n) =$$
$$= \frac{qD_p}{L_p} \frac{n_i^2}{N_D}\left[\exp\left(\frac{qV_J}{kT}\right)-1\right] \quad (20)$$

At high injection levels, the electron current, (electrons being majority carriers) in the n⁻ region of the junction is zero [3]:

$$J_n = q\mu_n n_n E + qD_n \frac{dn_n}{dx} \cong 0 \quad (21)$$

Drift-diffusion equilibrium is formed for electrons:

$$q\mu_n n_n E = -qD_n \frac{dn_n}{dx} \quad (22)$$

by the means of the electric field:

$$E = -\frac{kT}{q}\frac{1}{n_n}\frac{dn_n}{dx} \quad (23)$$

The hole current is accordingly given by:

$$J_p = q\mu_p p_n E - qD_p \frac{dp_n}{dx} = -qD_p\left(1+\frac{p_n}{n_n}\right)\frac{dp_n}{dx} \quad (24)$$

Since the n⁻ region of the junction is at high injection level ($n_n \cong p_n$), $J_p$ becomes:

$$J_p = -2qD_p \frac{dp_n}{dx} \quad (25)$$

Now the diffusion equation is obtained as:

$$2D_p \frac{d^2 p_n}{dx^2} - \frac{p_n - p_{n0}}{\tau_p} = 0 \quad (26)$$

and the solution will be also an exponential decrease, but with the decay constant $L_p' = \sqrt{2D_p \tau_p}$.

$$p_n(x) = p_{n0} + n_i \exp\left(\frac{qV_J}{2kT}\right)\exp\left(-\frac{x-x_n}{L_p'}\right) \quad (27)$$

This agrees well with eq.(9), since the injection level is high.
Similar to eq. (20), $J$ is computed as:

$$J = \frac{2qD_p}{L_p'} p_n(x_n) = \frac{\sqrt{2}qD_p}{L_p} n_i \exp\left(\frac{qV_J}{2kT}\right) \quad (28)$$

($V_J \gg kT/q$)

Since both (20) and (28) have similar dependence on $p_n(x_n)$, the general formula of the current-voltage characteristics $J(V_J)$ **regardless of the injection level** can be deduced from the general equation (15) for $p_n(x_n)$ as:

$$J = \frac{qD_p}{L_p}\left\{\sqrt{2n_i^2\left[\exp\left(\frac{qV_J}{kT}\right)-1\right]+N_D^2} - N_D\right\} \quad (29)$$

Is is easy to see that this equation reduces to (20) and (28), respectively, depending on the low/high injection level.

## 2.3. THE INFLUENCE OF THE SERIES RESISTANCE

At even higher current levels, the series resistance of the semiconductor material "bends" the exponential dependence of the current on voltage to a linear, ohmic dependence. Its influence can be included in the general formula (27) in the following manner:

$$J = \frac{qD_p}{L_p}\left\{\sqrt{2n_i^2\left[\exp\left(\frac{V_J - A_J JR}{kT/q}\right)-1\right]+N_D^2} - N_D\right\} \quad (29)$$

where $A_J$ is the junction area and $R$ is the series resistance. It is true that now $J$ dependence on the voltage is no longer explicit, but an explicit relation $V_J(J)$ can be written, since (29) has only one occurrence of $V_J$, unlike the case of $J$.